\documentclass{elsarticle}

\usepackage{mathart}

\usepackage{algorithm2e}

\usepackage{nicecolors}

\usepackage{xr}

\usepackage[numbers]{natbib}
\graphicspath{
  {figures/}
  {_data/plot/}
}

\title{
  Stochastic Epidemic Models inference and diagnosis
  with Poisson Random Measure Data Augmentation
}

\author[1,2]{Benjamin Nguyen-Van-Yen\corref{cor}}
\ead{bnguyen@biologie.ens.fr}

\author[3]{Pierre Del Moral}

\author[2,4,5]{Bernard Cazelles}

\address[1]{
 Institut Pasteur, Unité de Génétique Fonctionnelle des Maladies Infectieuses,
 UMR 2000 CNRS,
 Paris, France
}

\address[2]{
  Institut de Biologie de l'ENS (IBENS),
  Ecole Normale Supérieure, CNRS, INSERM, Université PSL,
  75005 Paris, France
}

\address[3]{
  INRIA, Bordeaux Research Center,
  France
}

\address[4]{
  International Center for Mathematical and Computational Modeling of Complex Systems (UMMISCO),
  UMI 209, UPMC/IRD,
  France
}

\address[5]{
 iGLOBE, UMI CNRS 3157,
 University of Arizona,
 Tucson, Arizona, United States of America
}

\cortext[cor]{Corresponding author}

\begin{document}


\maketitle

\section*{Abstract}

We present a new Bayesian inference method for compartmental models
that takes into account the intrinsic stochasticity of the process.
We show how to formulate a SIR-type Markov jump process
as the solution of a stochastic differential equation with respect
to a Poisson Random Measure (PRM),
and how to simulate the process trajectory deterministically from
a parameter value and a PRM realisation.

This forms the basis of our Data Augmented MCMC,
which consists in augmenting parameter space with the unobserved PRM value.
The resulting simple Metropolis-Hastings sampler acts as an efficient
simulation-based inference method, 
that can easily be transferred from model to model.

Compared with a recent Data Augmentation method based on Gibbs sampling of individual
infection histories, PRM-augmented MCMC scales much better
with epidemic size and is far more flexible.

PRM-augmented MCMC also yields \textit{a posteriori} estimates
of the PRM, that represent process stochasticity,
and which can be used to validate the model.
If the model is good, the posterior distribution
should exhibit no pattern and be close to the PRM prior distribution.
We illustrate this by fitting a non-seasonal model to some simulated
seasonal case count data.

Applied to the Zika epidemic of $2013$ in French Polynesia,
our approach shows that a simple SEIR model cannot correctly reproduce both
the initial sharp increase in the number of cases as well as the final
proportion of seropositive.

PRM-augmentation thus provides a coherent story for Stochastic Epidemic Model
inference, where explicitly inferring process stochasticity helps
with model validation.

\section{Introduction}

Stochasticity plays an important role in infectious disease dynamics,
but statistical inference of stochastic models is difficult.
When trying to estimate the parameters $\theta$ of
a complex stochastic epidemic model from some data $D$,
the data $D$ are typically an incomplete observation of the process,
and the observed data likelihood $\muP(D\ |\ \theta)$ is intractable
\cite{oneill_tutorial_2002}.
To work around this problem, one can decide to not use the likelihood
(ABC, deep learning),
to approximate the model (linear noise approximations),
to estimate the likelihood
(PMCMC \cite{andrieu_particle_2010,moral_alive_2015}),
or to rephrase the problem.

Data Augmentation (DA) introduces the alternative problem
of estimating the joint posterior $\muP(\theta,\ \nu\ |\ D)$,
instead of $\muP(\theta\ |\ D)$,
where $\nu$ is some latent variable chosen so that the joint likelihood
$\muP(D\ |\ \theta,\ \nu)$ is tractable.
The original problem then finds its solution by marginalizing over $\nu$,
$\muP(\theta\ |\ D) = \int \muP(\theta,\ \nu\ |\ D)\ d\nu$.
Data Augmentation has proven effective with MCMC
\cite{cauchemez_bayesian_2004,drummond_beast:_2007,jewell_bayesian_2009},
for dealing with granular, subject-level data, and complex models.

An important difficulty with DA is to explore the high dimensional
augmented state space effectively with the MCMC proposal.
Each class of models requires designing specific proposals,
and typically it is necessary to manually tune the proposal
for the specific problem at hand.
For instance in BEAST \cite{drummond_beast:_2007,bouckaert_beast_2014},
for which the latent state is the reconstructed phylogeny,
many different moves are implemented to explore tree space effectively,
and one is chosen at every iteration following certain probabilities.
For a given model and dataset, one should manually tune those probabilities
to achieve good mixing.

Additionally, it is challenging to design a proposal such that mixing speed
scales well with population size, which becomes necessary when dealing
with even moderately large epidemics.

We propose a new data augmentation scheme to fit simple models
to low granularity data -- like incidence or prevalence data.
This scheme is directly applicable
to the large class of Markov jump processes.

Pure jump Markov processes are widely used
for modelling infectious diseases,
ecological and evolutionary systems.
It is possible to represent them as solutions of Poisson-driven
stochastic differential equations,
which are integral equations with respect to a Poisson Random Measure (PRM).
This indicates a natural data augmentation
parametrization $(\theta, \nu)$ with $\theta$ the parameters of the process,
and $\nu$ the discrete measure that the process is integrated against.
A value of $\theta$ and a realisation $\nu$ of the PRM
are enough to deterministically simulate the trajectory $X$ of the process,
and so the joint likelihood of $\theta$ and $\nu$ is easily computed
as the probability of observing $D$ given the trajectory $X$.
Throughout the article, we will call this method PRM augmentation.

Different methods have already been proposed for the inference of
Markov jump processes, some specific to epidemic models
\cite{cauchemez_bayesian_2004,neal_case_2005,xiang_efficient_2014,neal_forward_2015,fintzi_efficient_2017},
and some more general 
\cite{rao_fast_2013,golightly_bayesian_2015,zhang_efficient_2017}.

In this article, we present PRM augmentation,
used with Metropolis-Hastings sampling
\cite{metropolis_equation_1953,hastings_monte_1970},
and discuss its advantages and drawbacks,
in terms of ease of use, speed, and insights,
and illustrate this on some synthetic and real datasets.

PRM augmentation relies on a generalization of uniformization
\cite{jensen_markoff_1953,rao_fast_2013,zhang_efficient_2017}
that is directly applicable to any Markov jump process,
with writing the Stochastic Differential Equation (SDE)
being the only mathematical work needed.
As such, it is a simulation-based method,
which can be used easily to compare different models.

We show how PRM augmentation compares to
a recent data augmentation method \cite{fintzi_efficient_2017},
and exhibits better complexity with respect to epidemic size.

Additionally, the natural distinction that is made between
the mechanistic part, estimated through $\theta$,
and the stochastic part, estimated through $\nu$,
provides a way to evaluate model fit and to propose model improvements,
similarly to bayesian latent residuals \cite{lau_new_2014},
\textit{via} the posterior $\nu$ estimates.

To demonstrate this, we apply the method to simulated data,
and to 2013-2014 Zika data from French Polynesia.

\section{Theory}

\subsection{
  Integration of a Markov pure jump process w.r.t. a Poisson Random Measure
}

Continuous time Markov chains, or Markov pure jump processes,
are stochastic processes whose state is constant in between jumps,
and jump only at exponential times, and that have the Markov property.
The trajectories of such processes are
right continuous, left limit, and piecewise constant.
A complete presentation can be found in \cite{ethier_markov_2009}.
Such processes are widely used in stochastic modelling,
and notably for instance in pomp \cite{king_statistical_2016}
or BEAST \cite{drummond_beast:_2007}.

Poisson random measures (PRM) are a generalization of Poisson processes
to more general spaces,
as random variables that take discrete measure values.
For a PRM, the number of points found in any measurable subset $A$
follows a Poisson distribution, and
the number of points found in two disjoint measurable subsets are independent.

A Markov pure jump process can be formulated as the solution
to a Poisson driven stochastic differential equation (SDE).
This is an established result \cite{ikeda_stochastic_1989},
and can be seen as a generalization
of uniformization \cite{jensen_markoff_1953}.
We provide a proof in the appendix \ref{app:proof} that solutions to 
the equation that interests us here are càdlàg and Markov,
and have the right infinitesimal generator.

The SDE hints at a natural deterministic algorithm to simulate
a trajectory $X$ of the process exactly,
given a realisation $\nu$ of the PRM.
Examine the points of $\nu$ ordered by time ;
for each successive point, compute the rate of the event corresponding
to that point ; if the value $u$ of the point is below the rate, 
then accept the event and update the state accordingly,
and if not reject the event (the state remains the same).
This iteratively draws a trajectory, with jumps that can only be
positioned at points of $\nu$, as illustrated in \fref{sim_exact_prm},
and following algorithm \ref{algo:exact}.

A value of $\theta$ and a value of $\nu$ uniquely determine the trajectory 
of the process, and thus are enough to compute the likelihood of observing
the data $\muP(D | \theta, \nu)$.

The a priori independence of $\theta$ and $\nu$ helps mixing
in the case of weakly informative data \cite{papaspiliopoulos_general_2007},
provided we are able to simulate the process efficiently.

\subsection{Simulation of a Markov Jump Process}

We consider a Markov Jump Process (MJP) $X$ with a state space $E = \Z^d$,
and a finite number of events $K$.
The $k$-th event happens with a rate $r_k : E \rightarrow \R_+$,
and when it happens, the state changes from $x$ to $x + \mu_k$.
Let $(\nu_k)_{k \leq K}$ be discrete measures on $\R_+ \times \R_+$.
\eref{integr} is the Poisson-driven SDE defining the process.

\begin{equation}
  \label{eq:integr}
  X_t = X_0
      + \sum_{k \leq K} \int_0^t \int_{\R_+}
        \one_{u \leq r_k(X_{s^-})} \mu_k\
        \nu_k(ds, du)
\end{equation}

This equation specifies a way to simulate the process.
Each point of the $(\nu_k)$ is a potential event point.
Consider all points of the $(\nu_k)$ ordered by time,
and for each one, compare the event rate to the $u$ coordinate of the point.
If the point is below the rate, then the event happens,
and we update the state of the system and the rates.
Otherwise nothing happens, and we continue on to the next point.
This yields the simplified algorithm \ref{algo:exact}.

\begin{algorithm}
  \SetKwInOut{Input}{input}\SetKwInOut{Output}{output}

  \Input{
    The discrete measures $(\nu_k)_{k \leq K}$,
    the rates $(r_k)$,
    the increments $(\mu_k)$,
    and the initial condition $X_0 \in E$
  }
  \Output{Trajectory $(X_t)_t$}

  Initialize $t_0 = 0$, $X_{t_0} = X_0$\;
  Iterate over the points of $(\nu)$ ordered by time,
  $(t_n,u_n,k_n)_n$\;

  \For{$n$ = $1$ \KwTo $\infty$}{
    \eIf{$u_n \leq r_{k_n}(X_{t_{n-1}})$}{
      $X_{t_n} = \mu_{k_n}(X_{t_{n-1}})$\;
    }
    {
      $X_{t_n} = X_{t_{n-1}}$\;
    }
  }

  \caption{Simplified algorithm for exact simulation of a MJP}
  \label{algo:exact}
\end{algorithm}

There is an obvious problem with algorithm \ref{algo:exact} however.
The $(\nu_k)$ have an infinite number of points on every time interval,
so we cannot order all of their points by time.
Instead we split $\R_+ \times \R_+$ into rectangles $A_{i,j}$,
and define $C_i = \bigcup_j A_{i,j} = [t_i, t_{i+1}]$ each time column.
In each rectangle $A_{i,j}$, the number of points $N^k_{i,j}$ of $\nu_k$
is finite almost surely,
and $\nu_k$ can be written
$\nu_k = \sum_{n \leq N^k_{i,j}} \delta_(t^{i,j,k}_n,u^{i,j,k}_n)$,
with the points $(t^{i,j,k}_n,u^{i,j,k}_n)$ ordered by time.
Equation \eref{integr:pieces} is equivalent to \eref{integr}

\begin{equation}
  \label{eq:integr:pieces}
  X_t = X_0
      + \sum_{k = 1}^K
        \sum_{i = 1}^\infty
        \sum_{j = 1}^\infty
        \sum_{n = 1}^{N^k_{i,j}}
        \one_{t^{i,j,k}_n \leq t}
        \one_{u_n \leq r_k(X_{{t^{i,j,k}_n}^-})}
        \mu_k
\end{equation}

We never need the rectangles that are above the rate.
We can use this fact to simulate the process with finite memory,
by lazily drawing points for the rectangles only when they are needed.
This exact algorithm is given in the appendix \ref{app:implem},
as algorithm~\ref{algo:app:exact}.

\begin{figure}
  \includegraphics[width=0.65\textwidth]
    {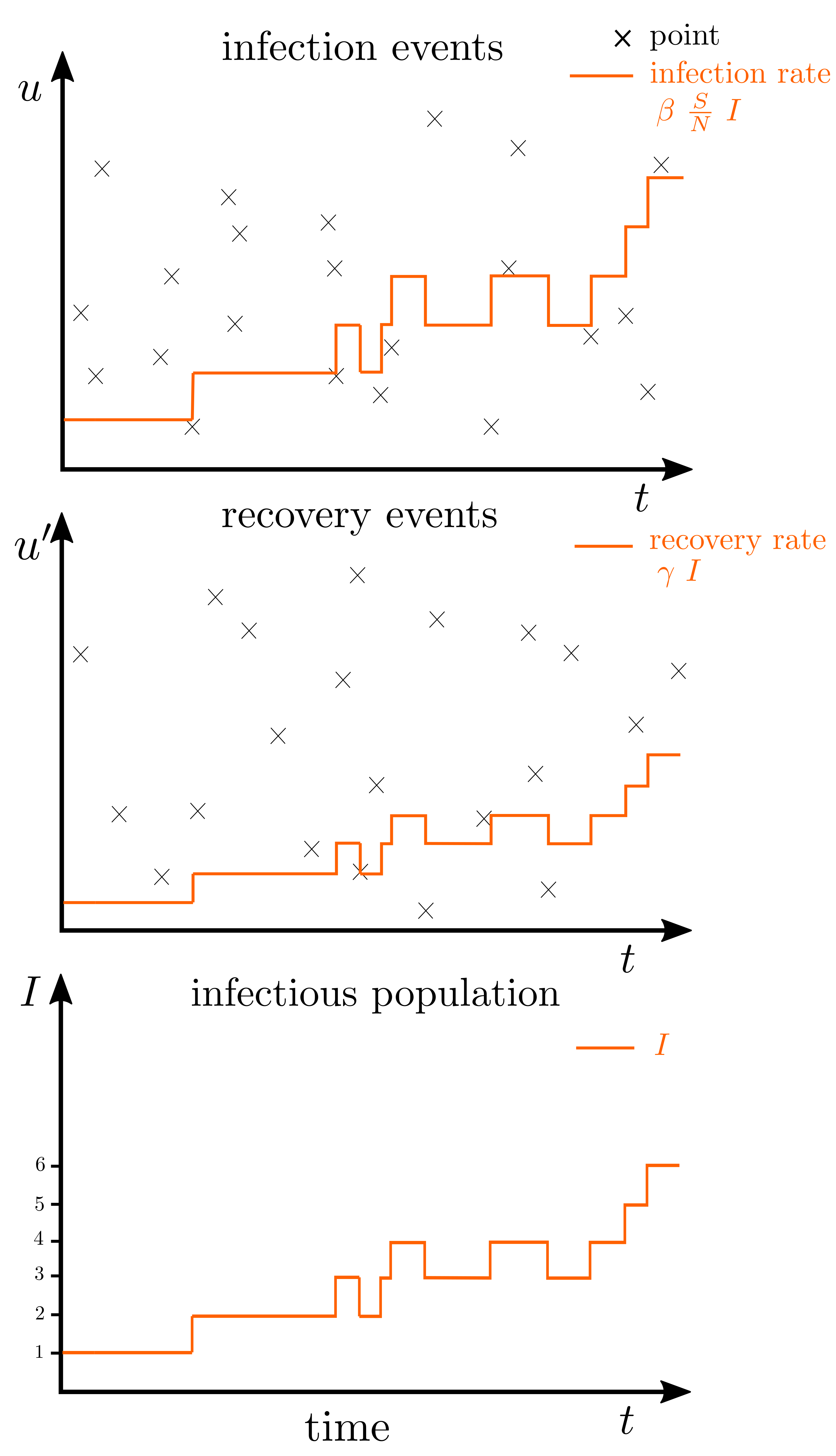}
  \caption[Exact simulation of a SIR process]
  {
    Exact simulation of a SIR process with respect to an infection and a recovery measure.

    The \black~crosses are the points of the infection and recovery measures
    respectively.
    The \orange~curves are the rates of infection and recovery respectively.

    When a point is under the rate curve, then the event happens.
    When an infection happens, $I$ (and the rates) increase,
    and when a recovery happens, $I$ (and the rates) decrease.
    Initially there is one infected and the whole population is susceptible.
  }
  \label{fig:sim_exact_prm}
\end{figure}

This exact algorithm is $O(n)$ in the number of points that are considered.
As a consequence, an important factor in its efficiency is
the proportion of points that are considered but rejected.
This is also true of classical uniformization, for which a constant
upper bound must be chosen in advance.
In contrast, here, the algorithm automatically adapts the upper bound 
to the variations of the rates,
so that the rejection rate can be kept small.

What's more, we can further reduce the complexity by
using an approximate simulation algorithm
$\tilde{X} = \tilde{f}(\theta, \nu)$.
To do that, we take inspiration from tau-leaping algorithms,
and treat the rates as if they were constant
on each time interval $[t_i, t_{i+1}]$,
$r_k^i = r_k(X_{t_i})$.
We then count the points below the rate $r_k^i$ in each interval.
We store the number of points present in each $A_{i,j}$ for each event
to make this faster.
The only points left to consider one by one are the ones
from the single $A_{i,j}$ such that $r_k^i$ is in $[u_j, u_{j+1}]$.
This approximate algorithm is written out in the appendix \ref{app:implem},
as algorithm~\ref{algo:app:approx}.

Once equipped with efficient simulations, we can use this PRM augmentation
scheme for parameter inference with MCMC.

\FloatBarrier

\subsection{PRM-augmented MCMC}

We want to estimate the posterior distribution of the parameter values
$\theta$ and of the discrete measure $\nu$ of the model
$(X_t) = f(\theta, \nu)$, given the data $D$, that is
\[
  \muP(\theta, \nu | D) \propto \muP(D | (X_t)) \muP(\theta) \muP(\nu) 
\]

$\muP(\theta)$ is the prior distribution placed on $\theta$,
and $\muP(\nu)$ is the PRM distribution.

We can use the Metropolis-Hastings algorithm
\cite{metropolis_equation_1953,hastings_monte_1970},
in the same way as for a deterministic system.
At every iteration of the MCMC, one draws new values $\theta'$ and $\nu'$
from the proposal, simulates the corresponding trajectory
with $f$ or $\tilde{f}$,
$(X_t)' = f(\theta', \nu')$,
computes the likelihood to observe the data $\muP(D | (X_t)')$,
then accepts or rejects the move according to the Metropolis-Hastings ratio.

For a PRM, points in disjoint subsets are independent.
Thus, if $\nu$ is a PRM sample, we can choose a subset of $\nu$,
erase its points and redraw new points from the PRM process,
and obtain a new (correlated) PRM sample.
This proposal $Q_\nu$ is reversible with respect to the PRM prior,
and thus can be used for MCMC.
More details are given in the appendix~\ref{app:mcmc}.
$Q_\nu$ can be used with any proposal $Q_\theta$ for the parameters,
with density $q_\theta$ to form a Metropolis-Hastings proposal,
as shown in algorithm \ref{algo:metropolis}.

\begin{algorithm}
  \SetKwInOut{Input}{input}\SetKwInOut{Output}{output}

  \Input{$\theta$, $\nu$}
  \Output{$\bar{\theta}$, $\bar{\nu}$}

  $\theta' \sim Q_\theta(.| \theta)$\;
  $\nu' \sim Q_\nu(. | \nu)$\;
  $X' = f(\theta', \nu')$ (or $X' = \tilde{f}(\theta', \nu')$)\;
  $\alpha =
    \frac{\muP(D | X')}{\muP(D | X)}
    \frac{\muP(\theta')}{\muP(\theta)}
    \frac{q(\theta|\theta')}{q(\theta'|\theta)}
  $\;
  $u \sim U([0, 1])$\;
  \eIf{$u < \alpha$}{
    \Return $X'$\;
  }{
    \Return $X$\;
  }

  \caption{PRM Metropolis-Hastings proposal}
  \label{algo:metropolis}
\end{algorithm}

\FloatBarrier

\subsection{Stochastic differential equation for a SIR model}

\label{sec:sir}

To illustrate our method, we show how to apply it to the classical SIR model
with case count data, and give both the deterministic \eref{sir:ode}
and stochastic \eref{sir:prm} formulations of the model here,
to show the correspondence between them.

$S$ is the number of susceptible hosts, $I$ the number of infected hosts,
$R$ the number of removed hosts,
and $C(t)$ the total number of cases, observed upon infection, up to time $t$.

$\beta$ is the effective contact rate,
and $\gamma$ is the rate of recovery (inverse duration of infection),
while $\rho$ is the reporting probability,
that is the probability that when an infection happens, we observe it.

\begin{equation}
  \label{eq:sir:ode}
  \begin{aligned}
    \frac{dS}{dt} &= - \beta \frac{S}{N} I \\
    \frac{dI}{dt} &= \beta \frac{S}{N} I - \gamma I \\
    \frac{dR}{dt} &= \gamma I\\
    \frac{dC}{dt} &= \rho \beta \frac{S}{N} I\\
                N &= S + I + R \\
  \end{aligned}
\end{equation}

To make the notations less heavy in the SDE,
we define the total rate of infection
$\lambda(s) = \beta \frac{S_s}{N_s} I_s$.
Also let $\nu^I$ and $\nu^R$ be independent PRMs on $\R_+ \times \R_+$
with intensity the Lebesgue measure,
for the events of infections and recoveries respectively.

The equations in \eref{sir:ode} and \eref{sir:prm} and their terms
are written in the same order so as to show the correspondence
between them.

\begin{equation}
  \label{eq:sir:prm}
  \begin{aligned}
    S_t &= S_0
         + \int_0^t \int_{\R_+}
            -~ \one_{u \leq \lambda(s^-)} \nu^I(ds, du)\\
    I_t &= I_0
         + \int_0^t \int_{\R_+} \left(
               \one_{u \leq \lambda(s^-)} \nu^I(ds, du)
           ~-~ \one_{u \leq \gamma I_{s^-}} \nu^R(ds, du)
         \right)\\
    R_t &= R_0
         + \int_0^t \int_{\R_+}
               \one_{u \leq \gamma I_{s^-}} \nu^R(ds, du)\\
    C_t &= \int_0^t \int_{\R_+}
             \one_{u \leq \rho \lambda(s^-)} \nu^I(ds, du)
         \\
    N_t &= S_t + I_t + R_t
  \end{aligned}
\end{equation}

We also use more complex versions of the SIR model,
by making the effective contact rate seasonal,
by adding host demography, immigration, immunity loss (SIRS) or a latent class
(SEIR).
To obtain the corresponding SDE, we only need to add a new PRM for every event,
and adapt the event rates.

For example, to add host demography,
we add the four events of birth of susceptible,
death of susceptible, death of infectious, and death of removed,
with rates $B N^\ast$, $D S$, $D I$, and $D R$,
and the four independent PRMs, $\nu^B$, $\nu^{DS}$, $\nu^{DI}$ respectively.

To make the model seasonal, we would replace $\beta$ by
$\beta(t) = \beta_m + \beta_v sin(2 \pi + \phi)$.

The equations for these different models are given in the appendix
\ref{app:results}.

To perform inference on the model, we need an observation model
to define the likelihood.
For incidence data, the observations are the number of cases
in each time interval $C(t_i) - C(t_{i-1})$,
and for inference we consider that the case count in each interval
is negative binomial.
The additional degree of freedom compared to the more natural Poisson
distribution helps the fitting on real data.
For prevalence data, we consider that the prevalence
observed at time $t_i$ follows the binomial distribution
$\cur{B}\left(I(t_i), \rho\right)$.
In both cases, the observations are independent conditionally on the
trajectory, and the full likelihood is simply the product of the likelihoods.

We will also apply PRM-augmented MCMC to the simple SIR model with no seasonality,
host vitality, or immigration, with prevalence data,
in which case the observations follow the binomial distribution
$\cur{B}(I(t_i), \rho)$.

\section{Materials and Methods}

\subsection{OCaml implementation}

The algorithm is implemented in the OCaml language
\cite{noauthor_ocaml_nodate},
and the project repository is at
\href{https://gitlab.com/bnguyenvanyen/ocamlecoevo}
     {https://gitlab.com/bnguyenvanyen/ocamlecoevo}

\subsection{Comparison of inference methods}

\label{sec:methods:compare}

For the comparisons, we simulate daily prevalence data
with the simple SIR model over a one month duration,
with a population size $N$ taking values
$500$, $1000$, $2000$, $4000$ and $8000$.
We estimate the base effective contact rate $\beta$,
and the initial conditions $S_0$, $I_0$ and $R_0$.

We start the MCMC from the target parameter values,
to not take into account differences in convergence speed between methods.

The different methods we compare are
the PRM-augmented MCMC with exact simulations,
the PRM-augmented MCMC with approximate simulations,
and a subject-level data augmentation MCMC with Gibbs sampling
presented in \cite{fintzi_efficient_2017}.

For the PRM-augmented MCMC, we first sample $\nu$ from its prior
for $100$ iterations,
then continue with our custom proposal for $2 000 000$ iterations.

For the Gibbs sampler we perform $10 000$ iterations.
Its moves in process space consist of redrawing
the infectious history of one subject, conditional on the data
and of the histories of the other subjects.
To maintain mixing as $N$ increases, we keep the number of histories
redrawn per iteration proportional to $N$
-- $20$ for $500$ and $320$ for $8000$.

As a measure of computational complexity for the different estimation
procedures, we compute the amount of computational time it takes
to obtain one effective sample from the posterior.

The multi-dimensional effective sample size \cite{vats_multivariate_2015} is defined as
\[
  {mESS}(n) = n \left( \frac{| \Lambda |}{| \Sigma |} \right)^{\frac{1}{p}}
\]
where $n$ is the number of samples, $p$ the number of dimensions estimated,
$| . |$ the determinant, $\Lambda$ the covariance structure of the posterior,
and $\Sigma$ the asymptotic covariance matrix of the Markov chain.
$\Lambda$ can be estimated  with the sample covariance matrix,
and $\Sigma$ with the batch means covariance matrix.

The space we explore is infinite dimensional so this definition
is not directly applicable.
We use the values the process takes at the datapoints $i$,
$X_i$ instead.
The trajectories of the simple SIR model live in the 2-simplex,
so for $m$ datapoints, the total dimension is then $p = 3 + 2 m$ for
$\beta$, the initial conditions and the $(X_i)$.

\subsection{SEIRS seasonal model simulation and inference}

\label{sec:methods:seasonal}

$5$ years of data are generated with a deterministic SEIRS model
with seasonality and vitality.
On infection, the hosts go through the Susceptible, Exposed, Infectious,
Removed, and then back to the Susceptible class.
The equations of the model and the parameter values used are given
in the appendix \ref{app:diagnosis}.
Parameter values are chosen so that the data displays
a clear periodic signal of period $1$ year, from the attractor of the system.
We also use a constant SEIRS model, that is, the same model,
but where $\beta(t)$ is constant ($\beta_m = 0$).

The data is then inferred with the stochastic seasonal model,
and with the stochastic constant model by PRM-augmented MCMC.

\subsection{French Polynesia Zika data}

\label{sec:methods:zika}

The French Polynesia Zika dataset is composed of weekly case counts
for Moorea island and of seroprevalence data from a cohort of $196$ people
from the archipelago.
The case count data has been made available in
\cite{mallet_bilan_2015,champagne_structure_2016},
and the seroprevalence data in \cite{aubry_zika_2017,champagne_structure_2016}.

We fit a SEIR model with immigration to the data with PRM-augmented MCMC.
The seroprevalence likelihood is binomial with probability the removed
proportion.

The details of the model and the prior distributions are given in
the appendix \ref{app:zika}.

The PRM-augmented MCMC is run for $200 000$ iterations for
convergence and adaptation of the covariance matrix
\cite{vihola_robust_2012},
then for $10^6$ iterations, from which $1000$ samples are kept.

\FloatBarrier

\section{Results}

\subsection{Efficiency of PRM-augmented MCMC}

\label{sec:compare}

\begin{figure}
  \centerline{
    \includegraphics{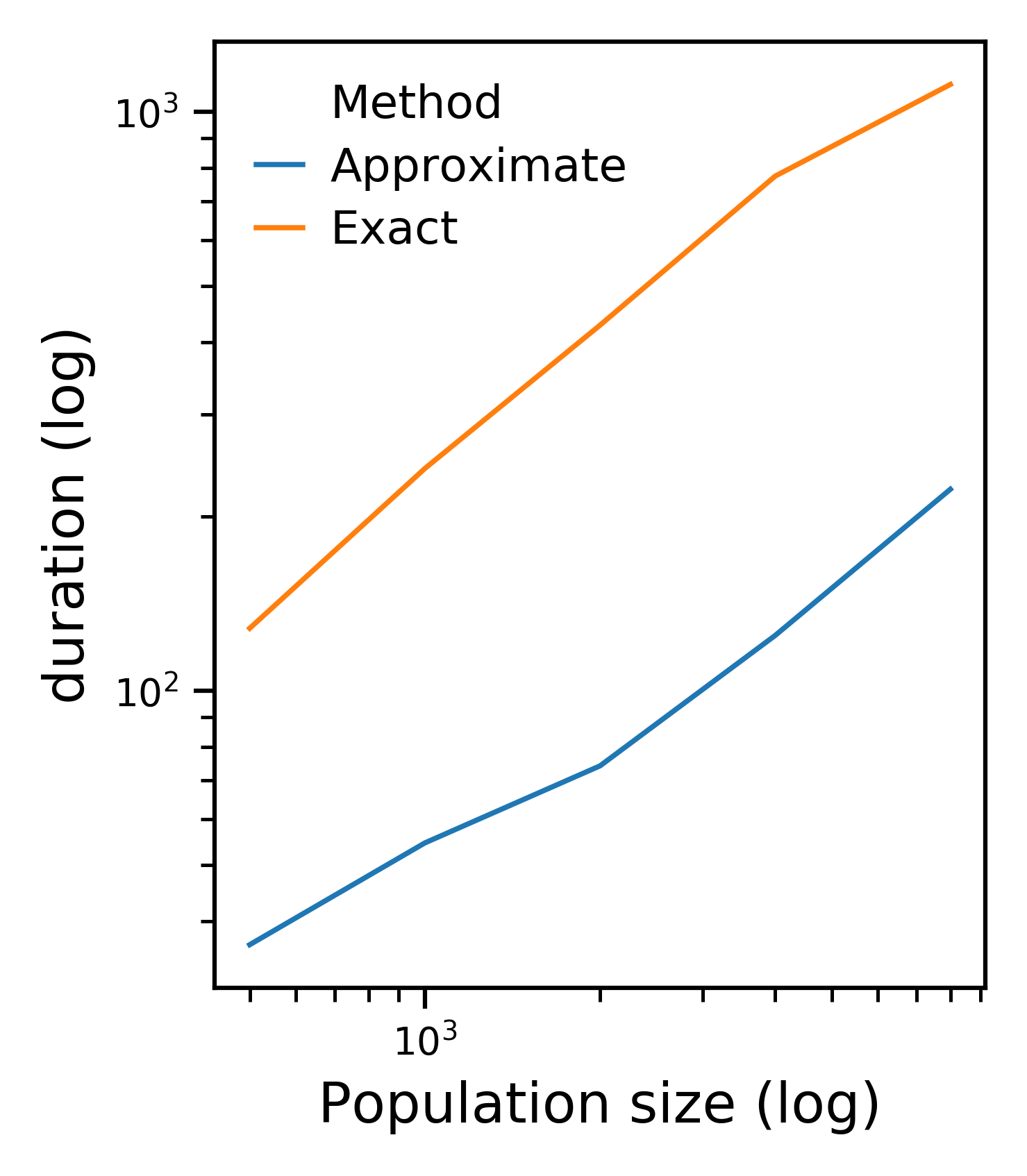}
  }
  \caption[Complexity of exact and approximate inference]
  {
    Comparison of execution time for exact and approximate simulations.

    Execution time (in minutes) as a function of host population size,
    on a log-log scale.
    In \blue, inference with the approximate simulations, and in \orange,
    with the exact simulations.
  }
  \label{fig:complex:exact}
\end{figure}

The comparison between exact and approximate simulations in
\fref{complex:exact} shows that both simulation schemes scale in the same order
on population size, but that the approximate scheme is much more economical.
In this example, for a population of $8000$, a million iterations
took a bit less than $4$ hours with the approximate method,
but close to $11$ hours with the exact method.
As population size increases, stochasticity plays a less important role,
and it becomes increasingly advantageous to use the approximate simulations.
For small populations and with high time resolution however,
it might be better to use exact simulations to avoid the bias
caused by approximation.

As both schemes rely on the same data structure for $\nu$,
it is easy to switch from one to the other,
and would be straightforward to implement switching when required
during estimation.

\begin{figure}
  \centerline{\includegraphics[width=\textwidth]
    {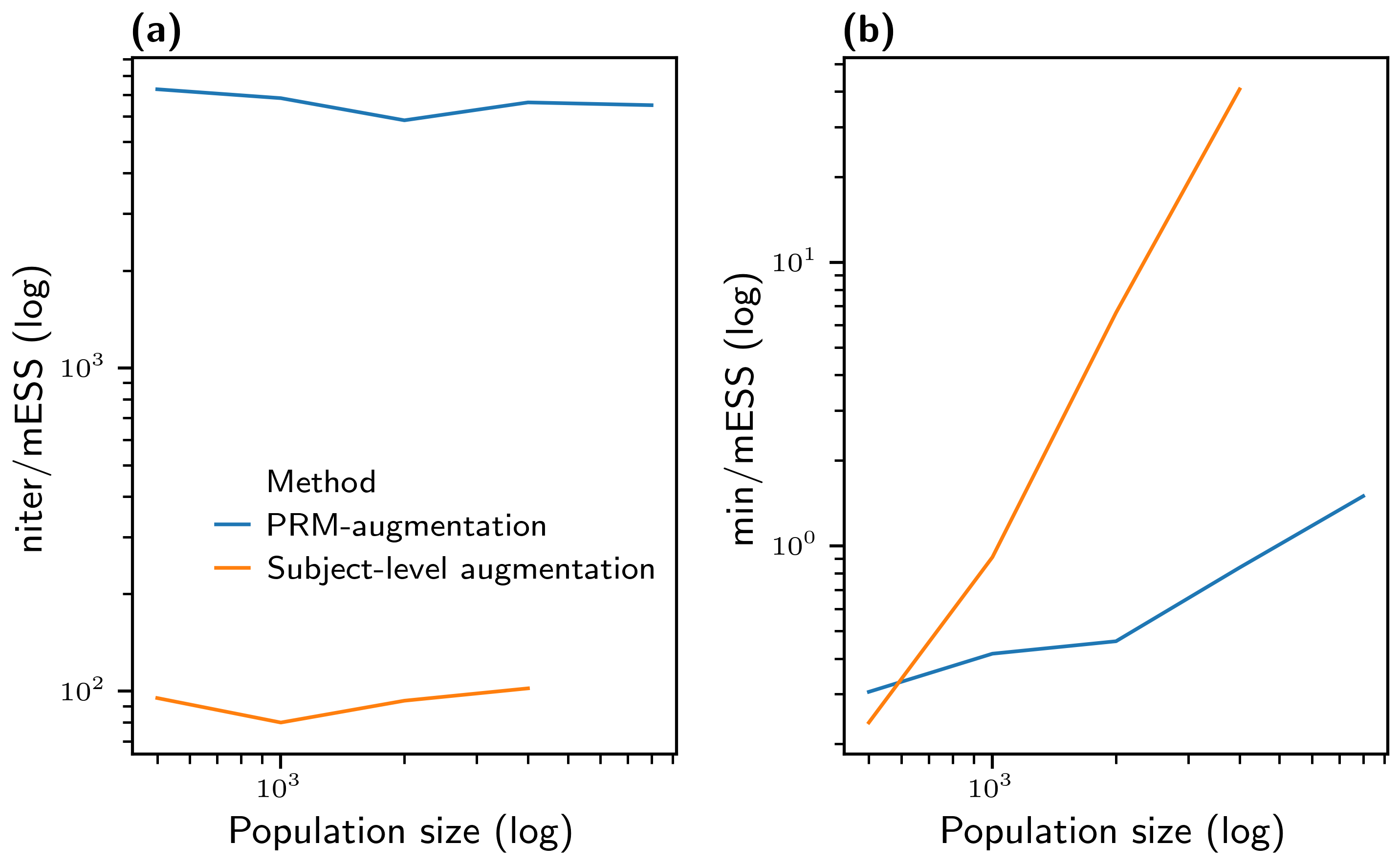}
  }
  \caption
    [Complexity of PRM-augmentation and subject level augmentation]
  {
    Comparison of complexity as a function of population size
    of PRM-augmentation and subject level augmentation.

    In \blue, PRM-augmentation,
    and in \orange, subject level augmentation.

    \textbf{(a)}
      Complexity measured as number of iterations ${niter}$ per effective sample

    \textbf{(b)}
      Complexity measured as CPU time in minutes per effective sample
  }
  \label{fig:benchmark}
\end{figure}

The PRM-augmented MCMC also proved much more efficient than
the subject level data augmentation method  described in
\cite{fintzi_efficient_2017}, as is shown in \fref{benchmark}.
The Gibbs sampler from \cite{fintzi_efficient_2017}
mixes much better than PRM-augmented MCMC when only looking
at the mESS \ref{sec:methods:compare} by iteration,
yielding around $1$ effective sample every $100$ iterations,
against $1$ every $6000$ iterations.
However, every one of its iterations is much more costly.
In theory, the complexity for an iteration of subject level augmentation
scales with the square of the population size.
Resampling a subject's history grows linearly with the number of events,
and the number of histories to redraw per iteration also grows linearly
with population size to maintain mixing.
This quickly becomes problematic, even with moderate population sizes.

In contrast, the mixing in the case of PRM-augmentation doesn't vary
much with population size, only the cost of an iteration does.
In practice, a linear regression indicates an exponent of $2.5$
for subject level augmentation, and $0.5$ for PRM-augmentation.

Example MCMC estimation for a population of $2000$ hosts is presented
in the Appendix \ref{app:compare} and \fref{app:benchmark:estim}.

\FloatBarrier

\subsection{Model diagnosis}

\label{sec:diagnosis}

To see the Markov jump process as the solution of an SDE
with respect to a PRM is to create a separation between the noise
driving the process, contained in the PRM, and the mechanism
of the process, described by the parameters of the process.
By inferring $\theta$ and $\nu$ jointly, then,
we are hoping to capture the mechanistic part in $\theta$,
and the noise part in $\nu$,
provided that the process explains the data well.

$\nu$ act as Bayesian latent residuals \cite{lau_new_2014}.
If the posterior distribution of $\nu$ is very different
from its prior distribution of a standard PRM (intensity one),
then our hypotheses about the noise in the model are wrong.
A pattern found in the posterior for $\nu$ means that
the pattern in the data is not entirely captured by the mechanism proposed,
or said another way, that the model is underfitting the data.
The nature of the pattern in the posterior for $\nu$
can then provide clues to improve the model to capture the pattern.

The posterior samples for $\nu$ can thus be used both
to evaluate model fit,
and to improve the model.

We illustrate this with an example on the SEIRS model,
in a seasonal and a constant version
(see \ref{sec:methods:seasonal}, page \pageref{sec:methods:seasonal}).
We generate some data with the seasonal model,
then compare the fit of this true seasonal model to the data,
to the fit of the constant model to the data.

We can see in $\fref{diagnosis}$, (a) that we are able to fit
the constant model to the seasonal data.
To evaluate the fit, we can look at the posterior samples of $\nu$.
We define the point density of $\nu$ for event $k$ and for a time slice $i$ as
the number of points by unit of volume,
$d^k_i = \frac{\nu(C^k_i)}{|C^k_i|}$.
The posterior $\nu$ estimates (\fref{diagnosis}, (b))
for the true seasonal model are very similar to the prior.
However for the constant model,
the infection point density rises at the start of every epidemic season,
and falls at the ends.
For the constant model to reproduce the data,
it is necessary to have more infections than expected
at the start of the epidemic season,
and less than expected at the end.
That is, the seasonal signal (\fref{diagnosis}, (c))
is captured in the posterior for $\nu$,
for us to notice and then include into the model.

\begin{figure}
  \centerline{\includegraphics[width=\textwidth]
    {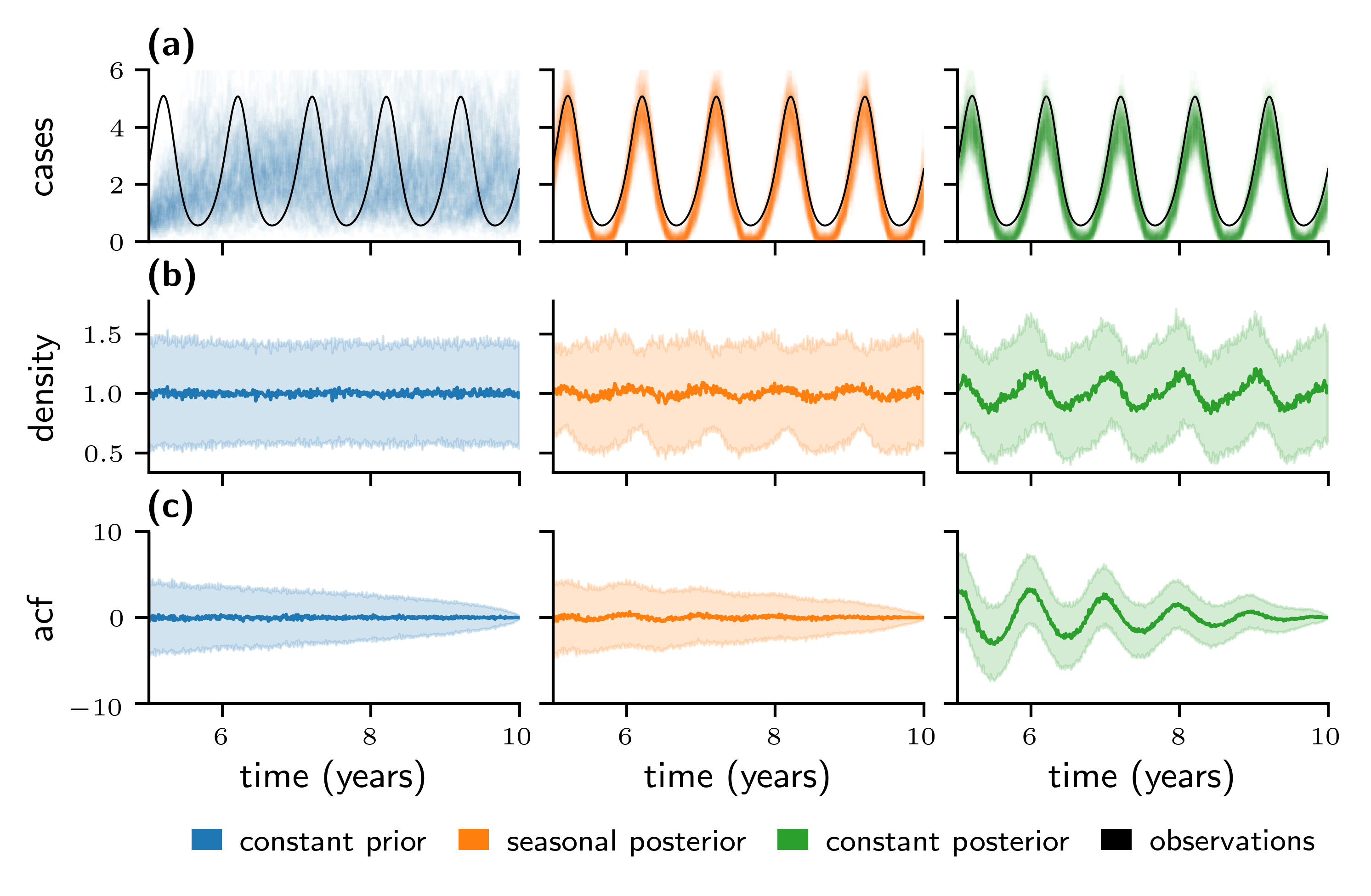}
  }

  \caption[Comparison of the seasonal and constant model fits]
  {
      Comparison of the fit of the seasonal and the constant model
      to simulated seasonal data.

      In \black, the simulated daily case count data.
      In \blue, in the first column,
      simulations with the true $\theta$ value,
      and $\nu$ sampled from its prior distribution.
      In \orange, in the second column,
      posterior samples for the seasonal model.
      In \green, in the last column,
      posterior samples for the constant model.

    \textbf{(a)}
      Daily case counts in the data (\black), and from MCMC samples.

    \textbf{(b)}
      Number of points by unit volume by slice of time
      for infection events
      in the posterior samples for $\nu$.
      The solid line is the mean, and the envelope the square deviation
      of the distribution.

    \textbf{(c)}
      Auto-correlation function of the function in (b)
      (Mean, and square deviation envelope).

      The average effective contact rate $\beta_m$,
      the initial conditions $(S_0,I_0,R_0)$,
      and the discrete measure $\nu$ are estimated.
      For both models, $250$ $\nu$ samples are kept for estimation,
      out of $500 000$ iterations after convergence.
  }
  \label{fig:diagnosis}
\end{figure}

This example is artificial, since the seasonal signal is already
very obvious in the data and would be included in the model
from the start.

However, it leads to two interesting remarks.
First, even though the constant model explains the seasonal data very badly,
we are still able to fit it to the seasonal data.
This shows that the MCMC is able to reach \textit{a priori} very unlikely
discrete measure values,
and so that we are able to explore the discrete measure space well.
Second, if we look at the estimates for the other events,
we can see that recovery is nearly the symmetric of infection.
The model doesn't really make a difference between more infections
and less recoveries,
or variation in infection rate, and variation in recovery rate.
This shows that this procedure is not a replacement for actually including
time-varying parameters \cite{cazelles_accounting_2018},
with a prior chosen to reflect how likely we expect those variations to be.

\subsection{French Polynesia Zika epidemic}

\label{sec:zika}

During the Zika epidemic in French Polynesia, in 2013-2014,
the case counts indicate that the epidemic progressed very fast,
but seroprevalence studies show that only around $50 \%$ of the population
got infected.

\begin{figure}
  \centerline{
    \includegraphics{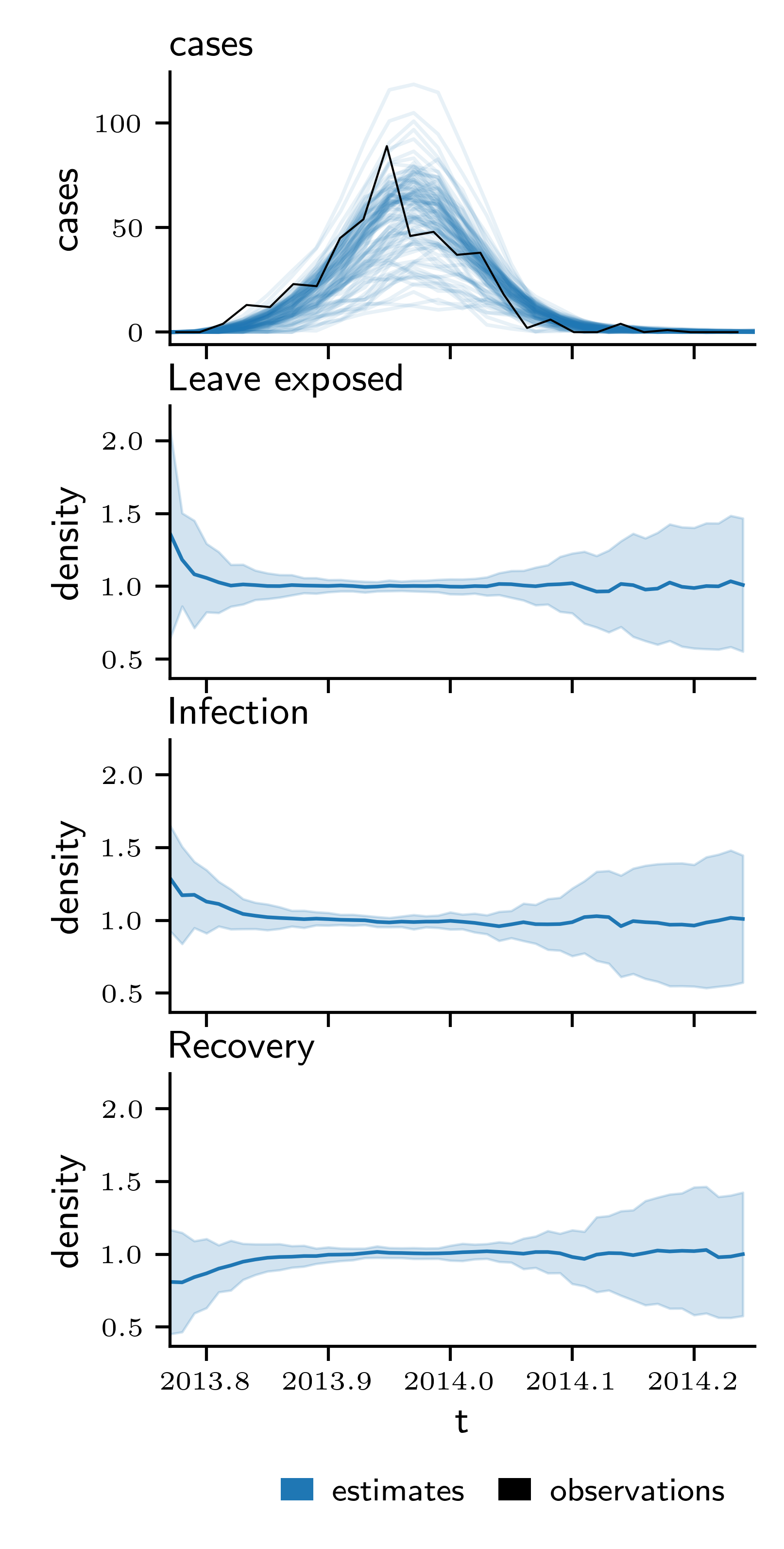}
  }
  \caption[French Polynesia Zika]
  {
    Fit of the SEIR model to the Zika data from Moorea island, 2013-2014.

    In \black, the data.
    In \blue, posterior samples for the stochastic model.

    The first panel represents the weekly number of cases, in the data,
    and in the posterior samples.

    The following panels represent the average and standard deviation
    of the density of points by unit of volume, in the $\nu$ posterior
    samples, for the events of infection, $S \rightarrow E$,
    leaving the exposed class, $E \rightarrow I$,
    and recovery, $I \rightarrow R$.

    $1000$ samples were kept from $10^6$ iterations after convergence.
  }
  \label{fig:moorea}
\end{figure}

The fit of a simple SEIR model with PRM-augmented MCMC shows that
the model cannot correctly capture the data,
if we restricut ourselves to realistic incubation and infectious periods,
as shown in \fref{moorea}.
The posterior $\nu$ distribution contains peaks of point density
at the beginning of the epidemic, most notably for the event of becoming
infectious, and for the event of infection.
This means that trajectories of the model that fit to the data
have more infections and shorter incubation periods
than expected by intrinsic noise,
only at the start of the epidemic.
There is no reason to expect that this could happen.
Said another way, in the initial period, 
cases happen faster than the model can follow.
More realistically, this could also be explained by an increase
in the reporting probability as the epidemic becomes noticed,
or also by population structure
\cite{champagne_structure_2016,kucharski_transmission_2016}.

\FloatBarrier

\section{Discussion}

Epidemiological data is ever more abundant and complex,
and can help us better understand the dynamics of infectious diseases,
answer important theoretical questions
and address public health problems.
Existing inference methods, however, are quickly becoming a limiting factor,
because of a prohibitive computational cost,
a difficulty in applying the method to arbitrary models,
or both.
The search for new methods progresses in different directions,
to build methods that are both faster and more easily transferable
to different models.

For MCMC, one such advance has been the development of adaptive methods
\cite{haario_adaptive_2001,roberts_examples_2009,vihola_robust_2012}.
They can relieve the user from the burden of designing proposals
taking into account the particular structure of the model,
as the adaptive method aims to discover that structure automatically.

The adoption of these methods for complex models takes time.
In the case of stochastic processes,
the problem is complicated by the very large (infinite dimensional)
latent variable space to explore.

As a consequence, a large part of the data augmentation methods
for stochastic processes tend to concern themselves with Gibbs sampling,
or Metropolis-Hastings component-wise updates
when Gibbs sampling cannot be achieved.

For Markov jump processes, an efficient sampler based on
uniformization \cite{jensen_markoff_1953}
was proposed in \cite{rao_fast_2013,zhang_efficient_2017}.
They make a clever use of uniformization to apply a method
from Markov chains to MJPs.
This method does not scale well to large state spaces however,
and so it is not practical in the case of stochastic epidemic models.

Instead, data augmentation methods targeted specifically
at stochastic epidemic models have also been developed these last 20 years
\cite{cauchemez_bayesian_2004,neal_case_2005,jewell_bayesian_2009,neal_forward_2015,fintzi_efficient_2017}.
They typically use component-wise updates, the effectiveness of which
depends crucially on the degree of posterior independence of the different
components.
In these methods, the latent variables represent the subject-level disease
histories.
This is a centered parametrization \cite{papaspiliopoulos_general_2007},
as the latent variables $\nu$ are a sufficient statistic for $\theta$.
By opposition, one can also use a non-centered parametrization,
in which the latent variables $\nu$ and $\theta$ are \textit{a priori} independent,
or phrased another way, $\nu$ is an ancillary statistic for $\theta$.

In the centered case, when the data brings little information,
$\theta$ and $\nu$ are very correlated and separate updates
for $\theta$ and $\nu$ must be very conservative.
In that case, non-centering can be much more effective,
as $\theta$ and $\nu$ will be nearly independent.
If instead the data is very informative, then the reverse is true,
and the centered parametrization will mix better than the non-centered one.

This question of centering and non-centering is crucial for mixing speed
in the case of component-wise updates, and has already been discussed
in the literature around stochastic epidemic models \cite{neal_case_2005}.

For typical stochastic epidemic models, it is often possible
to re-parametrize the model in terms of uniform random variates,
for instance, and thus to obtain a non-centered parametrization.
This can be the basis for non-centered simulation-based MCMC
\cite{neal_forward_2015}.
However, the parametrization is \textit{ad hoc} and must be found
by the modeller.
In constrast, we believe PRM-augmentation provides a canonical
non-centered parametrization,
and thus it is a good choice for low-granularity data like we have shown here.
For the case when the data is more informative,
it should be possible to design joint adaptive proposals
that could move reasonably far in trajectory space.

Indeed, the increased efficiency compared with the subject-level
data augmentation method from \cite{fintzi_efficient_2017}
can be in part attributed to this.
The subject-level augmentation is a centered parametrization,
and the cost of using Gibbs sampling to obtain new histories is that only
one history can be resampled at a time.
Another method of conditional resimulation that should scale better
with population size is however presented in
\cite{pooley_c._m._using_2015}.

PRM-augmented MCMC presents an interesting tradeoff
for moderately large population sizes, and a possible alternative
to linear noise approximations \cite{fintzi_linear_2020} or PMCMC \cite{andrieu_particle_2010}.
At the same time, the method is easily used in practice,
as it is akin to a simulation-based method,
that one can directly transpose to new models,
with no lengthy and error-prone mathematical derivations or implementation.

It would be interesting to make a quantitative comparison with PMCMC,
but it is difficult to do this in a fair manner,
as the efficiency of PMCMC crucially depends on the number of particles used.

However, we would like to argue that
the main advantage of PRM-augmentation is that it is
in a sense a natural parametrization of the model,
which makes our assumptions about the nature of the stochasticity very clear.
It leads to a clear separation between the process mechanism,
desribed by $\theta$, and the process noise, described by $\nu$,
that facilitates interpretation and is very useful for model diagnosis
(section \ref{sec:diagnosis}).
The procedure is identical to the one presented in
\cite{lau_new_2014,lau_model_2019},
with two differences.
First, the $\nu$ samples are obtained as part of the inference,
and not after.
That way, we don't need to be able to compute $\nu = f^{-1}(\theta, X)$
(and in our case, $f$ is not injective).
Second, and most importantly, we have a clear meaning for what the 
latent residuals $\nu$ represent, so that it is easy to interpret
deviations from the prior.

In conclusion, PRM-augmented MCMC is a practical method
for the inference of moderate epidemics where stochasticity
cannot be ignored.
As a simulation-based method, one can easily apply it to many
different models, and the PRM samples obtained make it possible
to diagnose, compare, and propose better models.

\section*{Declaration of Competing Interest}

The authors declare no conflict of interest.

\section*{Acknowledgements}

BN and BC have been supported by a grant from Agence Nationale de la Recherche
for the PANIC project
(Targeting PAthogen's NIChe:
a new approach for infectious diseases control in low-income countries:
ANR-14-CE02-0015-01).
BN is a doctoral student at the Centre de Recherche Interdisciplinaire,
in the Université de Paris.

\bibliography{../zotero_bibtex.bib}

\end{document}